\documentclass[dvips,a4paper]{jpconf}
\usepackage[dvips]{graphicx}
\usepackage[dvips]{graphics}
\usepackage{xcolor}
\usepackage{amssymb}

\newcommand{\fulldiamond}{\mbox{$\blacklozenge$}}
\newcommand{\fulltriangle}{\mbox{$\blacktriangle$}}

\begin{document}

\title{Evidence for Intra-Unit-Cell magnetic order in  the pseudo-gap state of high-$\rm T_c$ cuprates}

\author{Y. Sidis and  P. Bourges}
\address{Laboratoire L\'eon Brillouin, CEA-CNRS, CE-Saclay, 91191 Gif sur Yvette, France}
\ead{yvan.sidis@cea.fr}

\begin{abstract}

The existence of the mysterious pseudo-gap state in the phase diagram of copper oxide superconductors and its interplay with unconventional {\it d-wave} superconductivity has been a long standing issue for more than a decade. There is now a growing number of experimental indications that the pseudo-gap phase actually corresponds to a symmetry breaking state. In his theory for cuprates,  C. M. Varma proposes that the pseudo-gap is a new state of matter associated with the spontaneous appearance of circulating current loops within $\rm CuO_2$ unit cell. This intra-unit-cell order breaks time reversal symmetry, but preserves lattice translation invariance. Polarized elastic neutron scattering measurements provide evidence for an intra-unit-cell  magnetic order inside the pseudo-gap state. This order could be produced by the orbital-like magnetic moments induced by the circulating current loops. The magnetic order displays the same characteristic features in $\rm HgBa_2CuO_{4+\delta}$, $\rm YBa_2Cu_3O_{6+x}$ and $\rm Bi_2Sr_2CaCu_2O_{8+\delta}$  demonstrating that this genuine phase is ubiquitous of the pseudo-gap of high temperature copper oxide materials. We review the main properties characterizing this intra-unit-cell magnetic order and discuss its interplay or competition with other spin and charge instabilities.

\end{abstract}



\section{\label{Intro} Pseudo-gap state and circulating current loop theory}

In the phase diagram of high temperature superconducting copper oxides, various anomalous electronic properties have been reported in addition to the occurrence of unconventional {\it d-wave} superconductivity. In normal state, these materials exhibit non-Fermi liquid properties and enter a mysterious pseudo-gap regime, characterized by the loss of density of states on certain portions of the Fermi surface below a temperature $\rm T^{\star}$. A large variety of theoretical models has been designed to account for these exotic electronic properties and to shed light on their interplay with non conventional superconductivity. Among them, the pseudo-gap theory of C. M. Varma \cite{Varma97,Varma99,Varma06} proposes that the pseudo-gap state is a long range ordered state and that fluctuations of its order parameter control not only the non Fermi liquid properties, but also are likely to mediate the  {\it d-wave} superconducting pairing \cite{Aji10}. Derived from the 3-band Hubbard model,  this mean-field theory postulated that strong electronic interaction involving the Coulomb repulsion between copper and neighboring planar oxygens give rise to the spontaneous appearance of circulating current (CC) loops within $\rm CuO_2$ unit cell. The CC-loops appear by pairs, turning clockwise and anti-clockwise. This Intra-Unit-Cell (IUC) order breaks time reversal symmetry, but preserves lattice translation invariance. Beyond the mean-field theory proposed by C. M. Varma, the existence of a CC-loop order is challenged by other theoretical studies. The numerical analysis carried by Thomale and Greiter \cite{Greiter07,Greiter08} does not confirm their existence. The quantum variational Monte Carlo simulation of Weber and co-workers \cite{Weber}  questions the stability of the CC-loops  suggested by Varma's model, but suggests that a CC-loop order could become much more stable in a 5-band Hubbard model,  where apical oxygen orbitals are also included. In addition, the existence of CC-loops in ladder copper oxide materials is not fully excluded \cite{Gabay08,Gabay10,Scalapino}. Beyond pure theoretical considerations,  several experimental observations provide strong encouragement for models based CC-loop order in copper oxide materials.

\section{\label{polar} Evidence for an Intra-Unit-Cell magnetic order}

\begin{figure}[t]
\center
\includegraphics[width=8cm,angle=270]{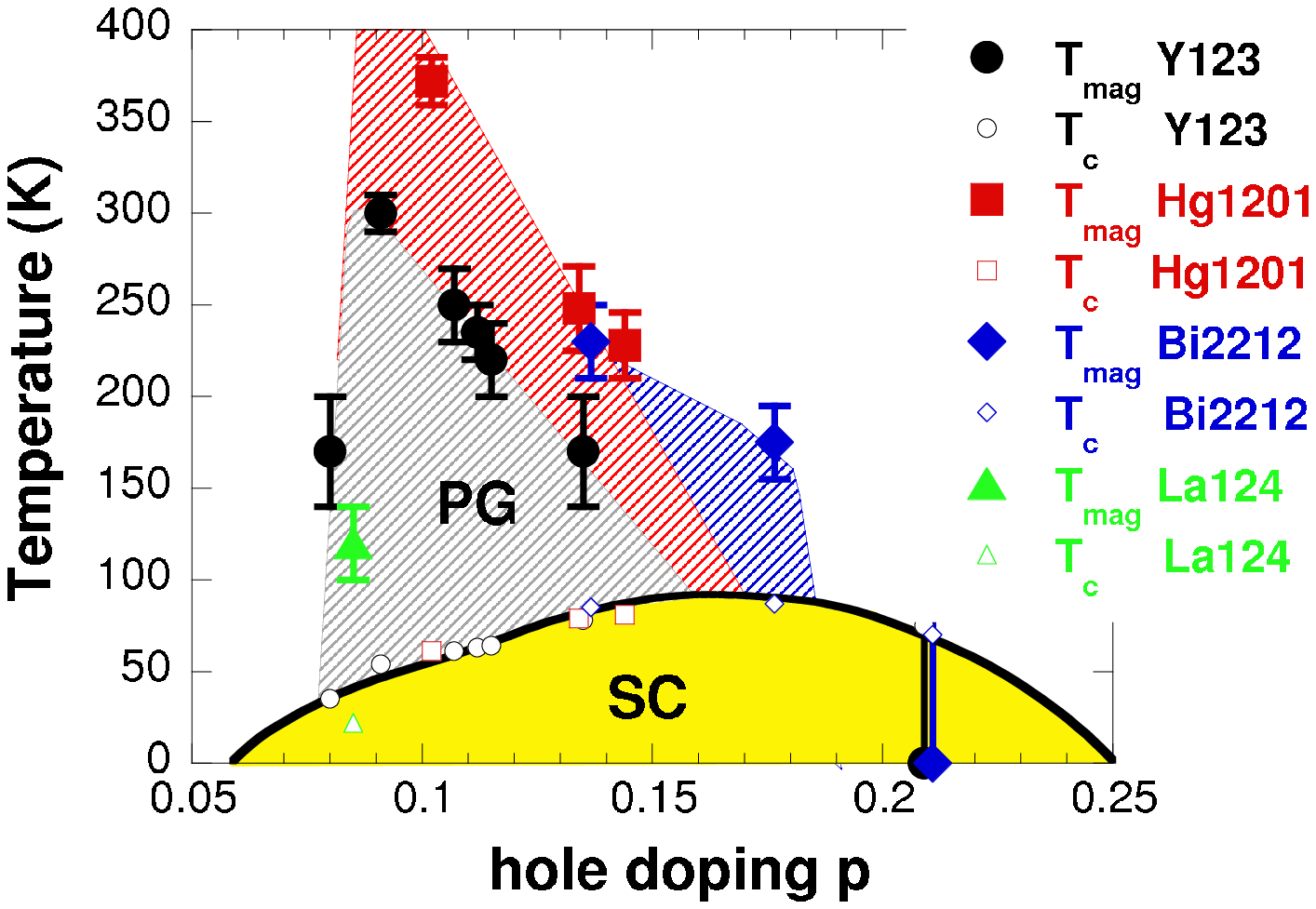}
\caption {(color online) Variation of the temperature $\rm T_{mag}$ at which the IUC magnetic order appears as a function the hole doping level for four cuprate families: $\rm YBa_2Cu_3O_{6+x}$ (Y123 - $ \color{black} \fullcircle $)) \cite{Fauque,Mook,Baledent-YBCO}, $\rm HgBa_2CuO_{4+\delta}$ (Hg1201- $\color{red} \fullsquare $) \cite{Li-Nature,Li-PRB}, $\rm Bi_2Sr_2CaCuO_{8+\delta}$ (Bi2212 - $\color{blue} \fulldiamond $) \cite{DeAlmeida-PRB}, $\rm La_{2-x}Sr_xCuO_4$ (La214 - $ \color{green} \fulltriangle$) \cite{Baledent-LSCO}. The open symbols stand for the value of the superconducting critical temperature $\rm T_c$ for each sample. Note that the value of $\rm T_c$ is a function not only of p but also of disorder, possibly leading to systematic differences in carrier concentration estimates. p for Hg1201 is estimated using the $\rm T_c(p)$ relationship for
Y123 \cite{Liang}.}
\label{Fig1}
\end{figure}

Polarized neutron scattering studies carried in bilayer $\rm YBa_2Cu_3O_{6+x}$ (Y123)  \cite{Fauque,Sidis,Mook,Baledent-YBCO} and monolayer $\rm HgBa_2CuO_{4+\delta}$ (Hg1201) \cite{Li-Nature,Li-PRB} have reported experimental evidence of a long range  3D magnetic order hidden in the pseudo-gap state. The existence of such a long range ordered magnetic state has been confirmed recently in bilayer $\rm Bi_2Sr_2CaCuO_{8+\delta}$ (Bi2212) \cite{DeAlmeida-PRB}. In  monolayer $\rm La_{2-x}Sr_xCuO_4$ (La214), a similar magnetic order has also been observed \cite{Baledent-LSCO}, but in this system it remains 2D and short ranged. Figure ~\ref{Fig1} shows the variation of the temperature $\rm T_{mag}$ at which the magnetic order settles in as a function of the hole doping level (p).

In all the three compounds with a superconducting critical temperature $\rm T_c$ as high as $\sim$90 K at optimal doping, the magnetic order develops concomitantly with the pseudo-gap state, at least for a hole doping level larger than p=0.09. In Hg1201 and Y123, the magnetic ordering temperature $\rm T_{mag}$ matches the pseudo-gap temperature $\rm T^{\star}$ determined by resistivity measurements \cite{Fauque,Li-Nature,CC-review}. In Bi2212 \cite{DeAlmeida-PRB}, the hole doping dependence and the magnitude of  $\rm T_{mag}$ and $\rm T^{\star}$  determined by various techniques \cite{Raffy,Ding,Ishida,STM} are quite consistent. The fact that $\rm T^{\star}$ does not decrease linearly with increasing hole doping and likely diplays a shoulder around optimal doping in Bi2212 \cite{Vishik} is well reproduced by the hole doping dependence of $\rm T_{mag}$ (Fig.~\ref{Fig1}). In addition, the comparison of  the polarized neutron scattering measurements and tunneling spectroscopy measurements in Bi2212 \cite{STM} suggest that the variation of the ordered magnetic moment upon increasing hole doping seems to reproduce the one of the pseudo-gap energy $\rm \Delta_{PG}$.

This  novel 3D magnetic order does not seem to depend strongly on the structural properties of the materials. In contrast with Y123 or Hg1201, Bi2212 has a body centered structure. The magnetic order seems to remain 3D independently from the nature of the stacking of Cu sites along the $\rm {\bf c}$ axis. Furthermore, Hg1201 is a tetragonal system, Y123 is orthorhombic along the Cu-O bonds and Bi2212 is orthorhombic along diagonals. The influence of orthorhombicity  on details of the magnetic structure is a relevant issue, but dedicated neutron scattering studies are still under progress.

Whatever the system where it has been observed, this novel magnetic state preserves the lattice translation invariant, but, at variance with ferromagnets, does not give rise to a uniform magnetization \cite{CC-review}. These observations imply the existence of an IUC (antiferro-)magnetic order. The overall symmetry of this order is consistent with the  so-called CC-$\rm \theta_{II}$ phase proposed by C. M. Varma \cite{Varma06}. Polarized neutron scattering studies may have detected the distribution of static magnetic fields generated by the CC-loops. The possible detection of these magnetic fields by local probes such as NMR, NQR or $\mu$SR is still debated \cite{Sonier-1,Sonier-2,Uemura,Strassle-2,Lederer}, but should help to get a deeper understanding of the intrinsic nature of the IUC magnetic order.

\begin{figure}[t]
\center
\includegraphics[width=12cm,angle=0]{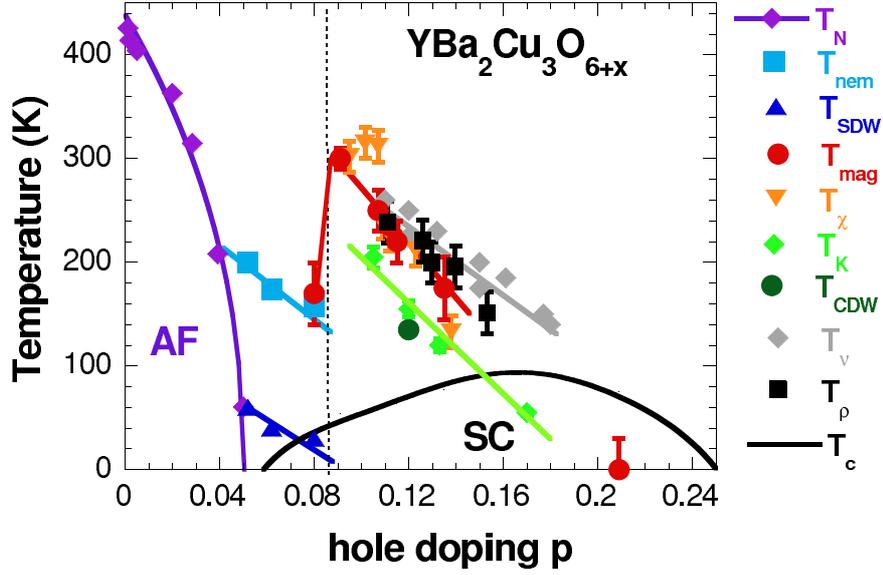}
\caption {(color online)  $\rm YBa_2Cu_3O_{6+x}$ phase diagram as a function of the hole doping p. In addition to the antiferromagnetic (AF) state and the d-wave superconducting (SC) state, several phases can be detected. In the lighly doped regime ($\rm 0.05<p<0.085$), incommensurate spin fluctuations develop a quasi-1D anisotropy spontaneously below $\rm T_{nem}$ \cite{Haug-review}. The $\rm C_4$ rotational symmetry is first broken, but the translation invariance is preserved. At lower temperature, below $\rm T_{SDW}$, the spin correlations freeze and a spin density (SDW) state  settles in \cite{Haug-review}. In the underdoped regime ($\rm p>0.085$), the pseudo-gap state develops below $\rm T^{\star}$ that can be determined by the departure of the electrical resistivity from a linear T-dependence at $\rm T_{\rho}$ \cite{Daou}.  The IUC-magnetic order appears at $\rm T_{mag} \sim T^{\star}$ \cite{Baledent-YBCO}. Around that temperature an anomaly can be observed in the second derivative of the magnetization at $\rm T_{\chi}$ \cite{Leridon} and the Nernst coefficient exhibits a net a-b anisotropy at $\rm T_{\nu}$ \cite{Daou}. Below $\rm T_{mag}$, a Kerr effect is observed at $\rm T_K$ \cite{Kapitulnik}. Around the specific hole doping $\rm p
\sim 1/8$, a charge density wave (CDW) instability develops below a temperature $\rm T_{CDW}$ \cite{Chang}, close to $\rm T_{K}$. }
\label{Fig3}
\end{figure}

Beyond polarized neutron scattering experiments, other measurements carried out in Y123 provide direct or indirect indications that a symmetry breaking state develops below the pseudo-gap temperature.
Indeed, resonant ultrasound spectroscopy measurements have recently reported the first thermodynamic evidence of a true phase transition taking place when entering the pseudo-gap \cite{Shekhter-sound}. In addition, the observation of anomalies in the  second derivative of the magnetization below $\rm T_{\chi}$ \cite{Leridon} can be understood in term of a coupling of the CC-loop order parameter to the uniform susceptibility \cite{Gronsleth}. Furthermore, the spontaneous appearance of an a-b anisotropy in the Nernst coefficient at $\rm T_{\nu}$ \cite{Daou} could be interpreted as a violation of the Onsager Reciprocity principle  in the CC-loop state which breaks time reversal symmetry \cite{Yakovenko}. Finally, a polar Kerr effect has been observed  within the pseudo-gap state below $\rm T_K$ \cite{Kapitulnik}. This effect has been initially interpreted as a signature of a time reversal symmetry broken state \cite{Kapitulnik}, but actually ascribed to a gyrotopic order \cite{Hosur}. As shown in Fig.\ref{Fig3}, the hole doping dependencies of the different temperatures $\rm T_{mag}$, $\rm T_{\chi}$, $\rm T_{K}$ and $\rm T_{\nu}$ display a certain similarity, suggesting that the related phenomena could share a common origin.

While the appearance of an IUC magnetic order can be ascribed to CC-loops, to date there is no direct observation of these CC-loops in superconducting cuprates. Alternatively, a resonant X-ray diffraction study in non superconducting CuO has claimed the first direct observation of CC-loops in $\rm CuO_2$ unit cell, i.e  the basic building block of superconducting copper oxide materials \cite{Scagnoli}, but has been challenged theoretically \cite{DiMatteo}. Nevertheless, such a kind of experiments in the superconducting cuprates would be very illuminating in resolving the question of whether
a CC-loop description is appropriate for the observed neutron scattering signal in the pseudo-gap phase \cite{DiMatteo}.

\section{\label{TRS} Time reversal symmetry breaking}

Since the discovery of the pseudo-gap regime, by pioneering NMR measurements in the early 90's \cite{Alloul-89}, the most accurate information concerning the electronic properties of this phase have been provided by  tunneling spectroscopy \cite{Renner-RMP} and angle resolved photo-emission (ARPES) \cite{Damascelli-RMP} in bilayer Bi2212. In the pseudo-gap state, left-circularly polarized photon give a different photocurrent from right-circularly photons in ARPES measurements \cite{Kaminski-Dichroism}. This circular dichroism would imply that time reversal symmetry is spontaneously broken in the pseudo-gap state, a phenomenon in good agreement with a CC-loop order \cite{Varma-Simon}.

\begin{figure}[t]
\center
\includegraphics[width=6cm,angle=270]{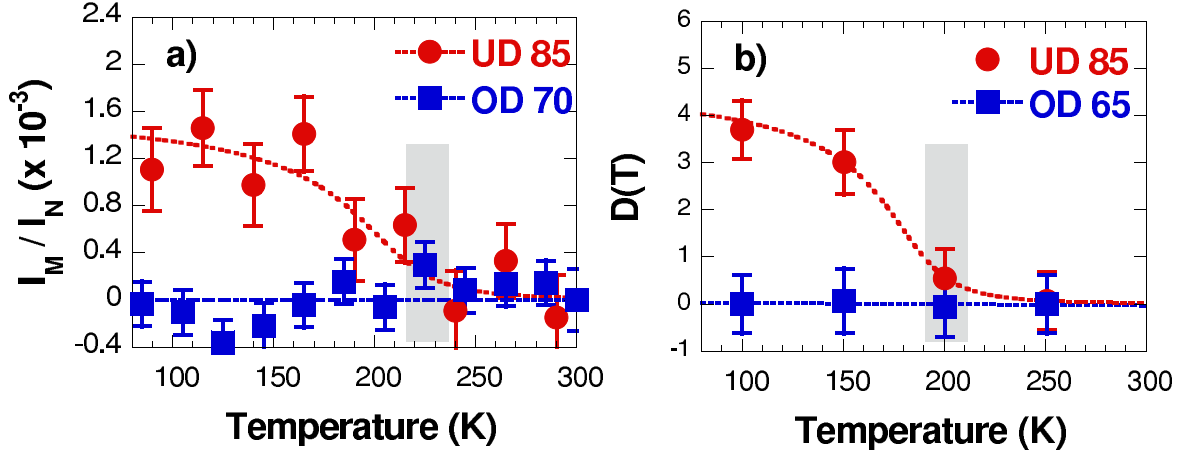}
\caption {(color online)  Comparison between polarized neutron scattering measurements performed on Bi2212 single crystals \cite{DeAlmeida-PRB} and circular dischroism obtained from ARPES measurements on Bi2212 thin films \cite{Kaminski-Dichroism}. a) The full magnetic intensity ($\rm I_M$) normalized by the nuclear intensity ($\rm I_N$) measured at the Bragg reflection (1,0,1). This normalization procedure is useful for comparing different samples. ($\rm I_M$) is about 3 orders of magnitude smaller than ($\rm I_N$). Raw polarized neutron scattering data and a detailed description of the method used to extract such a weak magnetic intensity have been recently given in \cite{DeAlmeida-PRB,CC-review}. b)  Circular dischroism D(T) at the anti-nodal wave vectors, reproduced from \cite{Kaminski-Dichroism}. The data associated with underdoped samples (UD- $\rm T_c$) are shown with symbols $\color{red} \fullcircle$ and those associated with overdoped samples (OD- $\rm T_c$)  with symbols $\color{blue} \fullsquare $.  }
\label{Fig2}
\end{figure}

The recent observation of an IUC magnetic order in Bi2212 further confirms that time reversal symmetry is broken in the pseudo-gap state.  The temperature dependencies of polarized neutron measurement in samples UD-85 and OD-70 \cite{DeAlmeida-PRB} and of the dichroic effects in samples UD-85 and OD-65 \cite{Kaminski-Dichroism} match with each other (Fig.~\ref{Fig2}). Keeping in mind that ARPES measurements are performed on thin films and polarized neutron measurement on large single crystals, the onset of time reversal  symmetry breaking found by both types of measurements for samples UD-85 is in a good agreement (Fig.~\ref{Fig2}). Likewise, the observations of dichroic effect at antinodal wave vectors and of a magnetic signal on Bragg reflections (1,0,L) indicate that the phase responsible for both effects possess the same symmetry as the CC-$\rm \theta_{II}$ phase proposed in the CC-loop theory for the pseudo-gap \cite{Varma-Simon}.

It is worth pointing out that the reported dichroic effect in ARPES and its interpretation are still controversial and the subject of a long running debate \cite{Borisenko,Norman,Arpiainen,Norman-comment,Lindros-reply}, since it has been argued that structural effects could also account for the dichroic signal reported in ARPES. Whenever correct for the analysis of ARPES measurements, these criticisms do not hold for polarized neutron diffraction measurements, since structural effects cannot produce a spin flip signal varying with the neutron spin polarization as observed experimentally.

Beyond circularly polarized ARPES measurements and polarized neutron diffraction, the observations of a polar Kerr effect at a temperature $\rm T_K$ in bilayer Y123 \cite{Kapitulnik} and in monolayer $\rm Pb_{0.55}Bi_{1.5}Sr_{1.6}La_{0.4}Cu_{6+\delta}$ (Pb-Bi2201) \cite{Pb-Bi2201} have been first presented as a direct evidence for time reversal symmetry breaking within the pseudo-gap state. This interpretation is now challenged by the fact the sign of the rotation of the polarization angle is the same on opposite surfaces of the sample, while in the usual Kerr effect the rotation angle must reverse.  While $\rm T_K$ coincides with $\rm T^{\star}$ in Pb-Bi2201, it appears deep inside the pseudo-gap state in Y123 (Fig.~\ref{Fig3}). Starting from a CC-loop order breaking time reversal symmetry at $\rm T^{\star}$, distinct theoretical models propose different routes to account for the observed polar Kerr once additional chiral properties are considered \cite{Varma-Kerr,Yakovenko-Kerr}.

All experimental reports available in monolayer systems (Hg1201, Pb-Bi2201) and bilayer systems (Y123, Bi2212) provide compelling evidence that the pseudo-gap phase is actually an ordered state, whose order parameter and broken symmetries remains to be clearly identified.

\section{\label{Tilt} Static and dynamical magnetic properties}

\begin{figure}[t]
\center
\includegraphics[width=8cm,angle=270]{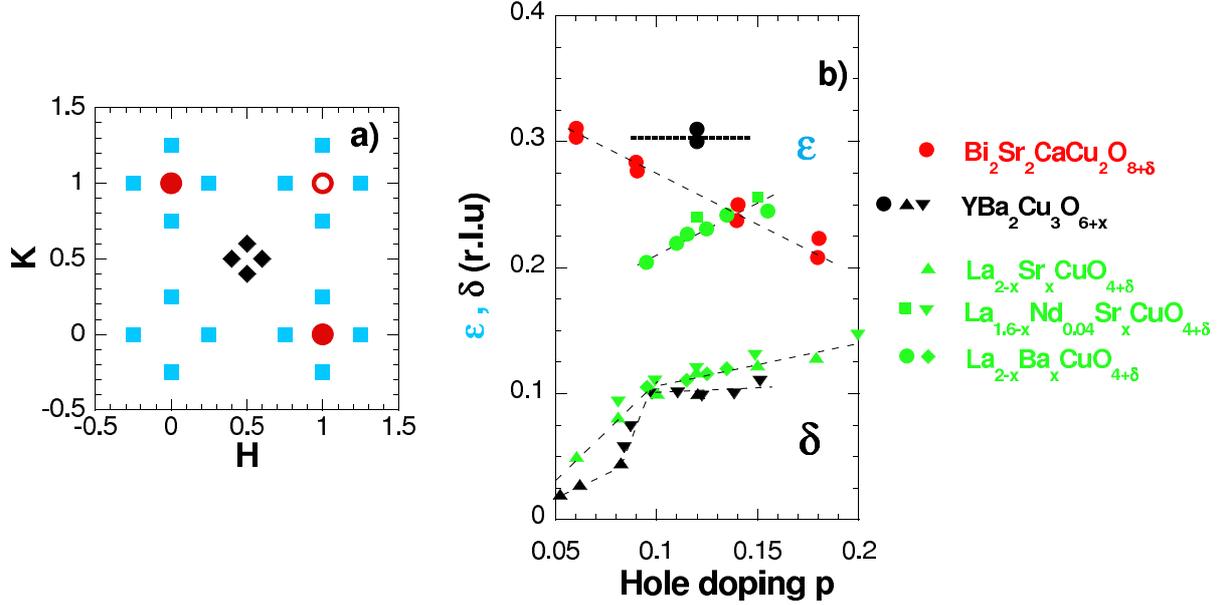}
\caption {(color online)  a) Location in the reciprocal space of the different kinds of spin, charge and orbital-like magnetic signals. The IUC-magnetic order is observed at the planar waves vector {\bf q }=(1,0) and (0,1) ($ \color{red} \fullcircle $), but should be observable at {\bf q }=(1,1) ($ \color{red} \opencircle $). A fluctuating or static charge density wave (CDW) is characterized by the planar wave vectors $\rm {\bf q}_{CDW}$=$\rm \pm (\epsilon,0)$ (and/or $\rm \pm (0,\epsilon)$ ($ \color{cyan} \fullsquare $). A fluctuating or static spin density wave (SDW) is associated with  the planar wave vectors $ \rm  {\bf q}_{SDW}$=$\rm  {\bf q}_{AF} \pm (\delta,0)$ (and/or $\rm \pm (0,\delta)$) ($ \color{black} \fulldiamond $), where  $\rm {\bf q}_{AF}$=(0.5,0.5). b) Hole doping dependencies of the incommensurability parameters $\rm \epsilon$ and $\rm \delta$ for 3 cuprate families:  (i) $\rm La_{2-x}Sr_xCuO_4$ \cite{Yamada98}, $\rm La_{1.6-x}Nd_{0.4}Sr_xCuO_4$ \cite{Tranquada-95,Tranquada-96,Tranquada-97,Tranquada-review,LNCO-X1,LNCO-X2}, $\rm La_{2-x}Ba_xCuO_4$ \cite{LBCO} (Light green symbols) ,(ii) $\rm YBa_2Cu_3O_{6+x} $ \cite{Haug-review,dai, Chang,Ghiringhelli}(black symbols), (iii) $\rm Bi_2Sr_2CaCu_2O_{8+\delta}$ \cite{Kohsaka} (red symbols). The parameter  $\rm \delta$ is measured using neutron scattering technique only, while $\rm \epsilon$ can be determined  using either by X-ray scattering or neutron scattering technique. For $\rm Bi_2Sr_2caCu_2O_{8+\delta}$, $\rm \epsilon$  is estimated from STM measurements.}
\label{Fig4}

\end{figure}

Neutron scattering technique can probe simultaneously the nuclear and magnetic correlations. Spin density wave (SDW) -like instabilities break translation invariance and produces a magnetic response at wave vectors distinct from the nuclear Bragg reflections. In cuprates, static or fluctuating SDWs are usually located around  the antiferromagnetic (AF) wave vector and  characterized by the wave vectors $ \rm  {\bf q}_{SDW}$=$\rm  {\bf q}_{AF} \pm (\delta,0)$ (and/or $\rm \pm (0,\delta)$) \cite{Tranquada-review}, where  $\rm {\bf q}_{AF}$=(0.5,0.5) stands for the planar AF wave vector, given in reduced lattice units (tetragonal notation) (Fig.~\ref{Fig4}.a). At variance, the IUC magnetic order implies the existence of staggered magnetic moments within the unit cell and  corresponds to a q=0  magnetic instability.  Since the IUC magnetic order preserves the lattice translation invariance, the weak magnetic signal that it produces in a neutron scattering experiment is superimposed with the much stronger nuclear scattering. The use of spin polarized neutron scattering technique is then essential to disentangle the nuclear and magnetic neutron scattered intensities. A full polarization analysis allows a determination of the orientation of the ordered magnetic moments.

The search for a long range 3D magnetic order in the pseudo-gap phase has been performed on Bragg reflections (1,0,L)-(0,1,L) with integer L values in monolayer  Hg1201 \cite{Li-Nature,Li-PRB} and in bilayers Y123 \cite{Fauque,Sidis,Mook,Baledent-YBCO} and Bi2212  \cite{DeAlmeida-PRB}.  In the case of monolayer La214 \cite{Baledent-LSCO},  the IUC order is quasi-2D and can be observed for any L value. In Ref.~\cite{DeAlmeida-PRB}, the magnetic intensities along the $\rm {\bf c}$ axis have been compared for these 4 cuprates families.  Interestingly,  these intensities  exhibit for all systems  a similar decay with $\rm \frac{2 \pi}{c} L$ (the component of the wave vector perpendicular to the $\rm CuO_2$ plane),  suggesting a common magnetic origin. For all cuprate families, the polarization analysis demonstrates that the magnetic moments cannot be strictly perpendicular to the $\rm CuO_2$ planes.  A  conservative estimate of  $\rm  45 \pm 20 ^{o}$ can be given for the tilt angle of these moments  with respect to the $\rm \bf {c}$ axis. While Polar Kerr effect is usually associated with the appearance of ferromagnetic moments, it can also be mediated by magneto-electric coupling for an antiferromagnetic order. The tilt of magnetic moments could be an important feature which lowers the symmetry of the IUC magnetic structure and allows the magnetic structure to be active to produce a Kerr effet \cite{Orenstein}.

The origin of the static magnetic order reported by polarized neutron diffraction is still an open issue. The IUC-cell order could be induced by CC-loops in the CC-$\rm \theta_{II}$ phase. This phase is characterized by two CC-loops per $\rm CuO_2$ unit cell and the loop pattern is fourfold degenerated. However, the CC-loops are confined with the $\rm CuO_2$ planes, generating orbital magnetic moments perpendicular to the planes. This is at variance with the experimental observation. Recently, two Ising magnetic mode has been discovered in the pseudo-gap state of Hg1201 \cite{Li-INS-1,Li-INS-2}. These new modes could be ascribed to magnetic excitations associated with a change of configuration of the CC-loop pattern. The two excitations corresponding to a $\pm$90$\rm ^{o}$ rotation of a loop pattern are indeed active in the spin flip channel  in inelastic polarized neutron scattering. The quantitative analysis of the characteristic mode energies and their weak dispersion leads to the conclusion that the ground state cannot reduce to one of the four states of the CC-$\rm \theta_{II}$ phase,  but should be made of their quantum superposition \cite{Varma-INS-1,Varma-INS-2}. This produces a quantum interference phenomenon in neutron diffraction measurement: the tilt angle of the magnetic moment reported in polarized neutron scattering measurements would then highlight the degree of admixture of the four orthogonal states of the CC-$\rm \theta_{II}$ phase.

Alternatively, it has been proposed that CC-loops could be delocalized on the $\rm CuO_6$ octahedra for a monolayer like Hg1201 \cite{Weber} or on $\rm CuO_5$ pyramids in bilayer systems  like Y123 or Bi2212 \cite{Lederer}. This model should produce a magnetic signal on Bragg reflection (00L), but no signal has been observed at Bragg position (002) neither in Y123 \cite{Fauque} nor in Hg1201 \cite{Li-Nature}. Furthermore, this model fails to account for the polarization analysis carried out on the Bragg reflection (100) \cite{CC-review}.

\section{\label{SDW} Competing spin instabilities}

In La214  \cite{Baledent-LSCO} for a hole doping level p=0.085, the quasi-2D and short range IUC magnetic order appears at a  rather low temperature $\rm T_{mag} \sim $120 K. Below this temperature, the dynamical spin response at the wave vector $\rm  {\bf q}_{SDW}$ exhibits two marked modifications: (i) an enhancement of  the low energy spin fluctuations intensity  and (ii) an order parameter-like increase of the incommensurability parameter $\delta$. Both effects indicate a deep interplay between the q=0 magnetic order and the incommensurate spin fluctuations.
In lightly doped Y123 \cite{Baledent-YBCO} (p $\sim$ 0.08), the 3D long range IUC order  develops  at $\rm T_{mag} \sim $170 K (Fig.~\ref{Fig1}), i.e very close to the temperature $\rm T_{nem} \sim $ 150 K (Fig.\ref{Fig3}). At this characteristic temperature, the low energy incommensurate  spin fluctuations start exhibiting a net a-b anisotropy  \cite{Haug-review}. $\rm T_{nem} $ is usually interpreted as the onset temperature at which the $\rm C_4$ rotational symmetry is spontaneously broken by strong electronic correlations pushing the system into an electronic nematic state. In lightly doped Y123, the orthorhombic lattice distortion serves as weak orientational field and is essential for the observation of the a-b anisotropy. Since this state also preserves the translation invariance, it corresponds to another competing q=0 electronic instability. At much lower temperature, spin correlations freeze at $\rm T_{SDW}$ (Fig.~\ref{Fig3}), giving rise to a spin density wave (SDW) state that can further coexist with the superconducting phase \cite{Haug-review}.
In underdoped Y123  \cite{Baledent-YBCO} ($\rm p > 0.09$), the 3D  IUC magnetic order is well developed and settles in at $\rm T_{mag} \simeq T^{\star}$. In parallel, low energy spin fluctuations around the AF wave vector are gapped. Through the substitution of a few percents of non magnetic Zn impurities (known to strongly weaken the superducting order),  neither  $\rm T_{mag}$ \cite{Baledent-YBCO} nor  $\rm  T^{\star}$ \cite{Alloul-review} are affected. On the contrary, the average magnetic moment associated with the IUC magnetic order  drops down. This  indicates that the IUC order is likely to be destroyed around Zn impurities, but remains unaffected far away. Simultaneously, quasi-1D incommensurate spin fluctuations are restored at low energy, suggesting that a competing SDW order could develop where the IUC magnetic order is suppressed. All these studies suggest a competition between the IUC magnetic state and an incipient SDW state.

\section{\label{CDW} Competing charge instabilities}

The existence of an IUC order associated with the pseudo-gap state in Bi2212 system is not only supported by polarized neutron scattering measurements. In Bi2212, the analysis of STM images at $\rm \Delta_{PG}$, the energy usually associated with the pseudo-gap,  points towards an IUC order in the pseudo-gap state \cite{Lawler}. At $\rm \Delta_{PG}$, the unbalance of the electronic densities measured at planar wave vector $\rm {\bf q}_x$=$\rm(1,0)$ and $\rm {\bf q}_y$=$\rm (0,1)$ suggests that $\rm C_4$ rotational symmetry is broken in the pseudo-gap state. The analysis of spectroscopic imaging STM is consistent with an IUC electronic nematic order, yielding distinct electronic density on O sites along $\rm{\bf a}$ and $\rm{\bf b}$ directions \cite{Lawler}.  In principle, the IUC electronic nematicity is different from the IUC loop order.  However, the mean-field analysis of the different IUC-ordering possibilities in the 3-band Emery model indicates that the electronic nematic order and the CC-loop order could actually coexist \cite{Fischer}.

The IUC electronic nematicity inferred from spectroscopic imaging STM studies corresponds to a long range order which coexists  with a (short range) charge density wave (CDW) order \cite{Lawler}. This order break the $\rm C_4$ rotational symmetry  and the lattice translation  invariance. The CDW correlations are characterized by the planar wave vectors $\rm {\bf q}_{CDW}$=$\rm \pm (\epsilon,0)$ (and/or $\rm \pm (0,\epsilon)$) (Fig.~\ref{Fig4}.a). According to STM measurements in Bi2212, the incommensurability parameter $\rm \epsilon$ is typically of the order of 0.25 r.l.u and decreases linearly with increasing hole doping \cite{Kohsaka} (Fig.~\ref{Fig4}.b). Furthermore, the fluctuating CDW order develops below the onset of the pseudo-gap state and can be viewed as a consequence of the pseudo-gap rather than its cause \cite{Yazdani}. Likewise, a local CDW order has been recently associated with topological defects in the electronic nematic state \cite{Mesaros}.

In Y123, the observation of an a-b anisotropy in the Nernst coefficient \cite{Daou} close to $\rm T^{\star}$ (Fig~\ref{Fig3}) is often interpreted as an indication that the $\rm C_4$ rotation symmetry is spontaneously broken within the pseudo-gap state and associated with an electronic nematic state \cite{Hackl}. At lower temperature, hard X-ray  \cite{Chang} and soft X-ray resonant scattering \cite{Ghiringhelli} studies reported evidence for a charge modulation along wave vectors $\rm {\bf q}_{CDW}$=$\rm \pm (\epsilon,0)$ and $\rm \pm (0,\epsilon)$, with $\epsilon \sim$ 0.3 r.l.u (Fig.\ref{Fig4}.b). Considering X-ray measurements, one cannot yet unambiguously  distinguish  between  an equal distribution of domains of two uniaxial CDWs with mutually perpendicular propagating wave vectors or a single CDW with biaxial charge modulation. However recent sound velocity measurements strongly support a single CDW with biaxial charge modulation \cite{Leboeuf}, but is at variance with  $\rm C_4$ rotation symmetry breaking inferred from Nernst coefficient. The intensity of the CDW signal grows on cooling down to $\rm T_c$  below which it is suppressed. Under an external magnetic field, the CDW is strengthened while superconductivity is depressed.  NMR measurements \cite{Julien} further indicate that a static  CDW  shows up only under magnetic field, but should be absent at zero field. As a consequence the CDW  reported in X-ray measurements should  be still fluctuating. In Y123, the evidence of a CDW instability is well documented in a narrow hole doping range around p $\sim$ 1/8 ($\rm 0.09 < p <0.13$) (Fig.~\ref{Fig4}.b, solid line). The CDW instability develops at $\rm T_{CDW}$ well below the pseudo-gap temperature $\rm T^{\star}$ or the temperature associated with the IUC magnetic order $\rm T_{mag}$ (Fig.~\ref{Fig3}).

In La214, there is no direct observation of a CDW order neither by neutron scattering nor by X-ray scattering. At variance, once the system is co-doped with Nd ($\rm La_{1.6}Nd_{0.04}Sr_xCuO_4$) \cite{Tranquada-95,Tranquada-96,Tranquada-97,LNCO-X1,LNCO-X2}  or Sr replaced by Ba ($\rm La_{2-x}Ba_{x}CuO_4$) \cite{LBCO}, a CDW order can be observed below the LTO-LTT structural transition and followed by a SDW order. In that specific cuprate family, $\epsilon$ is about twice $\delta$, suggesting a strong interplay between charge and spin modulations. These combined charge and spin modulations are usually interpreted within stripe models where doped hole segregate into charge stripes separating hole poor AF domains (\cite{Tranquada-review} and references therein). The stripe order develops around the characteristic hole doping $\rm p\sim 1/8$, where stripes could be pinned down onto the lattice. In La214 (p=0.085) \cite{Baledent-LSCO}, where there is neither indication for a CDW order nor for a SDW order, the IUC magnetic order remains quasi-2D and at short range.  The planar magnetic correlation lengths are consistent with a picture where CC-loops could be confined within 2-leg copper ladders. In this scenario, the CC-Loop order would be frustrated by the competing stripe instability, which seems much stronger in that cuprate family than in any other superconducting cuprate materials.

More theoretical and experimental work is needed to shed light on the interplay between the pseudo-gap phenomenon, the IUC orders (i.e q=0 instabilities) and the spin and/or charge density wave instabilities. To this respect, one can notice that for decades the role of oxygen in cuprates has been overlooked, whereas an IUC magnetic order could be bound  to CC-loops \cite{Varma97,Varma99,Varma06} or orbital oxygen moments \cite{Moskvin}. In addition, the observed charge modulations could be produced by a valence bond state or quadrupolar density wave state \cite{Pepin}, related to a non uniform electronic densities on oxygen sites. Thus, the internal degrees of freedom of $\rm CuO_2$ plaquettes could become relevant to shed light on the physics of cuprates.

\section{\label{SC} Superconducting state}

While polarized neutron scattering measurements show the existence of an IUC magnetic order in the normal state below $\rm T^{\star}$, the evolution of this order upon entering the superconducting state is still an open and interesting issue. Indeed time reversal symmetry is violated in this state, whereas it is preserved in a pure {\it d-wave} superconducting state \cite{Leridon-pwave}. However the search for indications of the competition or coexistence between these distinct electronic states using polarized neutron scattering technique has been limited by a technical problem. To date, the  elastic polarized neutron scattering measurements have been performed using a standard longitudinal polarization set-up \cite{CC-review}, for which the neutron polarization is maintained on the sample using a tiny magnetic guide field ($\sim$ 10 Gauss).  Since a depolarization of the neutron beam can take place when cooling down below $\rm T_c$, most of polarized neutron scattering studies have been restricted to the normal state. In early studies carried out in Y123 \cite{Fauque}, a few data in the superconducting state were reported, but no definitive conclusion can be drawn from these data, since the issue of depolarization of the neutron beam was not addressed. To overcome such a technical problem, measurements in the superconducting state require to place the sample in a zero magnetic field chamber, already available in a more sophisticate spherical polarization set-up (such Cryo-PAD or Mu-PAD). Using this device, elastic polarized neutron measurements will be safely extended into the superconducting state in a close future.

\section{\label{conclusion} Conclusion}

Polarized elastic neutron scattering measurements indicate that an  IUC magnetic order develops in the pseudo-gap state. This novel magnetic order displays the same characteristic features in monolayer Hg1201 and bilayer Y123 and Bi2212, demonstrating that this genuine phase is ubiquitous in the pseudo-gap of high temperature copper oxide materials. The polarized neutron scattering and the circularly polarized photoemission out over various cuprate families all point towards a breaking of time reversal symmetry up on entering the pseudo-gap state. The intrinsic nature of the IUC magnetic order remains to be understood, but, to date, the few information we have concerning the magnetic pattern within the $\rm CuO_2$ unit cell are consistent with the CC-loop model (phase CC-$\rm \theta_{II}$) proposed by C. M. Varma \cite{Varma97,Varma99,Varma06}. Nevertheless, the stability of the CC-loop state is still an open theoretical issue, in addition to the ability of such a q=0 instability to produce a gap in the charge excitation spectrum. New theoretical and experimental developments are needed to shed light on the interplay between the observed IUC magnetic order and other electronic states, such as an IUC electronic nematicity, incipient CDW/SDW orders and {\it d-wave} superconductivity.


\end{document}